\documentclass[acmsmall,screen,nonacm, ,bookmarksnumbered,unicode,hidelinks]{acmart}
\usepackage[labelfont=bf,textfont={bf}]{caption}
\usepackage{textcomp}
\usepackage{dblfloatfix}  
\usepackage{float}
\usepackage{balance}
\usepackage{placeins}
\usepackage{minted}
\usepackage{algorithm}      
\usepackage{algpseudocode}
\usepackage{graphicx} 
\usepackage{subcaption}
\usepackage{xcolor}
\usepackage{alltt}
\usepackage{url} 
\usepackage{rotating}
\usepackage{adjustbox}
\usepackage{booktabs}
\usepackage{enumitem}
\usepackage{algorithm}
\usepackage{lineno}
\usepackage{mdframed} 
\usepackage{fontawesome5}  
\usepackage{graphicx} 
\usepackage{array}    
\newcolumntype{C}[1]{>{\centering\arraybackslash}p{#1}}
\usepackage[table]{xcolor} 
\usepackage{multirow}
\usepackage[most]{tcolorbox} 
\usepackage{wrapfig}
\setitemize{noitemsep,topsep=0pt,parsep=0pt,partopsep=0pt}
\newcommand{\bi}{\begin{itemize}}
\newcommand{\ei}{\end{itemize}}

\newlength{\unitlengthforbar}
\definecolor{grayline}{gray}{0.8}
\definecolor{lightyellow}{RGB}{255,255,204}
\definecolor{lightpurple}{RGB}{229,204,255}
\definecolor{lightred}{RGB}{255,204,204}
\definecolor{lightgrey}{RGB}{224,224,224}
\definecolor{lightblack}{RGB}{82,82,82} 
\definecolor{lightblue}{rgb}{0.85,0.92,1}
\definecolor{lightgreen}{rgb}{0.85,0.95,0.85}
\definecolor{lightcyan}{rgb}{0.85,0.95,0.95}
\definecolor{lightorange}{rgb}{1.0,0.9,0.8}
\definecolor{lightpink}{rgb}{1.0,0.85,0.9}
\definecolor{gold}{RGB}{255,215,0}
\definecolor{skyblue}{RGB}{135,206,235}
\definecolor{lightgreen}{RGB}{144,238,144}
\definecolor{vlg}{RGB}{242,242,242}
\definecolor{lightpeach}{RGB}{250,230,215}
\newenvironment{conclusion}[1]
{
  \begin{mdframed}[
    leftline=true,               
    topline=false,               
    bottomline=false,            
    rightline=false,             
    linecolor=black,             
    linewidth=4pt,               
    backgroundcolor=gray!20,     
    innertopmargin=6pt,          
    innerbottommargin=6pt,       
    skipabove=10pt,              
    skipbelow=10pt,              
    innerleftmargin=6pt,        
    rightline=false            
  ]
  \textbf{#1} 
  \ignorespaces
}
{\end{mdframed}
}  

\usepackage{rotating}

 \newcounter{myquotecounter}

\usepackage{listings}
\usepackage{xcolor}
 
\lstdefinelanguage{pseudocode}{
  keywords={function, end, if, then, else, while, for, do, Input, Output, while, return},
  comment=[l]{//},
  basicstyle=\ttfamily,
  keywordstyle=\BLUE\bfseries,
  stringstyle=\color{red},
  commentstyle=\color{gray},
}

\lstdefinestyle{pseudocode}{
    language=pseudocode, 
    basicstyle=\ttfamily\scriptsize,
    keywordstyle=\BLUE\bfseries,
    identifierstyle=\color{black},
    commentstyle=\color{gray}\ttfamily,
    stringstyle=\color{red},
    showstringspaces=false,
    columns=fullflexible,
    numbers=none
}

\definecolor{niceblue}{HTML}{0000FF}

\newcommand{\there}[1]{%
  \hyperlink{resp:#1}{%
    \fcolorbox{black}{black!15}{%
      \bfseries\scriptsize #1%
    }~on page \pageref{resp:#1}%
  }%
}

\newcommand{\here}[1]{%
  \hypertarget{resp:#1}{}
  \fcolorbox{niceblue}{niceblue!15}{%
    \label{resp:#1}%
    \bfseries\scriptsize  {#1}%
  }%
}
\newcommand{\BLUE}{\color{blue}}
\renewcommand{\there}[1]{} 
\renewcommand{\here}[1]{}
\renewcommand{\BLUE}{\color{black}}

\begin{document}
\title{How Low Can You Go? The   Data-Light SE Challenge }
\author{Kishan Kumar Ganguly}
\email{kgangul@ncsu.edu}

\author{Tim Menzies}
\email{timm@ieee.org}

\affiliation{%
  \department{Department of Computer Science}
  \institution{North Carolina State University}
  \city{Raleigh}
  \state{NC}
  \country{USA}
}

\begin{abstract}
\vskip 0.1cm
Much of Software Engineering (SE) research assumes that progress depends on massive datasets and CPU-intensive optimizers. Yet has this assumption been rigorously tested? 

\here{A1} 
\BLUE
The counter-evidence presented in this paper suggests otherwise. For over 100 optimization tasks from recent SE papers
(including software configuration,  performance tuning, product line engineering, project health forecasting, defect prediction, software testing, software process and cost estimation, and cross-domain generalization datasets), even with just a few dozen labels, very simple methods (e.g., diversity sampling, a minimal Bayesian learner, its distance-based non-parametric variant, or random probes) achieve over $90\%$ of the best reported results. Furthermore, these simple methods perform just as well as more complex state-of-the-the-art optimizers like SMAC, TPE, DEHB etc. While some tasks would require better outcomes and more sampling, these results seen after a few dozen samples would suffice for many engineering needs (particularly when the goal is rapid and cost-efficient guidance rather than slow and exhaustive optimization). 

\color{black}
To say that another ways,  at least some SE tasks are better served by lightweight approaches that demand fewer labels and far less computation. 
We hence propose the     data-light challenge: when will a handful of  labels suffice for SE tasks? To enable a large-scale investigation of this issue, we contribute (1) a mathematical formalization of labeling, (2) lightweight baseline algorithms, and (3) results on public-domain data showing the conditions under which lightweight methods excel or fail. 

For the purposes of open science,   our scripts and data are online at \url{https://github.com/KKGanguly/NEO}.

\end{abstract}


\maketitle

\vskip 0.1cm
\section{Introduction}
Do all AI tasks in SE require   large and
  complex models?
We ask this since transformer-based models (with their trillion-parameter complexity) dominate much of the current literature, 
Such complexity is sometimes necessary.
 For example,
 the attention mechanisms of LLMs (that adjust conditional probabilities across context windows) need access to underlying models connecting trillions of variables~\cite{vaswani2017attention}.  
  
But do all SE tasks need such complexity?
  Perhaps not.
 When prioritizing massive test suites, researchers find that 2–3 variable models select failure-prone tests with high probability~\cite{ling2021different}. In search-based SE, multi-generational evolutionary methods barely outperform basic feature selection~\cite{menzies2007business,nair2016accidental}. And, as shown by the experiments of this paper,
 for 120+ SE optimization
 tasks, models built from just a few labelled examples   can match full-data performance~\cite{menzies2008implications,peters2015lace2}.
\here{A2}\BLUE Table~\ref{combinedtable} on the next page shows all the multi-objective optimization tasks  used in this study. 
These tasks  include   configuration tuning (e.g., during compiler optimization), as well as  project health forecasting (e.g. Github issue lifetime classification), product line engineering (e.g. high-dimensional constraint satisfaction problems for Scrum feature models), defect prediction and mitigation (e.g. ranking test features that find most bugs), and software process and cost estimation (e.g. regression dataset involving agile process simulation to estimate effort and risk). 
\color{black}

\begin{table*}[!t]
\scriptsize
\renewcommand{\baselinestretch}{0.9}
\caption{\here{A3}\BLUE 127 SE optimization problems from recent SE papers (stored in the MOOT repository).  “x/y’’ denotes inputs/outputs (independent/dependent attributes).}
\label{combinedtable}
\begingroup
\scriptsize
\renewcommand{\arraystretch}{0.55}  

\begin{adjustbox}{max width=\textwidth}
\begin{tabular}{@{}C{1.4cm}@{~}p{2.5cm}p{3.9cm}p{4.0cm}p{1.1cm}p{1.4cm}p{2.8cm}@{}}

\# Datasets & Dataset Type                     & File Names                                                               & Primary Objective                                                     & x/y          & \# Rows & Cited By       \\ \midrule
25          & \begin{tabular}[c]{@{}l@{}}Specific Software\\Configurations\end{tabular} & SS-A to SS-X, billing10k                                                 & Optimize software system settings                                     & 3-88/2-3   & 197–86,059 & \cite{Amiraliminimaldata, menzies2025the, lustossa2024isneak, senthilkumar2024can, lusstosa2025less, nairMSR18, peng2023veer,nair2017using}   \\ 
12          & \begin{tabular}[c]{@{}l@{}}PromiseTune Software\\Configurations\end{tabular} & \begin{tabular}[c]{@{}l@{}} 7z, BDBC, HSQLDB, LLVM, PostgreSQL, \\ dconvert, deeparch, exastencils, javagc, \\ redis, storm, x264\end{tabular}                                                 & Software performance optimization                                     &  9-35/1  &  864-166,975 &  \cite{chen2026promisetune, chen2025accuracy, DBLP:conf/icse/XiangChen26, DBLP:journals/pacmse/Gong024, gong2024dividable} \\ \midrule
1           & Cloud                            & HSMGP num                                                                & Hazardous Software, Management Program data                           & 14/1         & 3,457 &  \cite{Amiraliminimaldata, menzies2025the, chen2025accuracy, senthilkumar2024can}     \\
1           & Cloud                            & Apache AllMeasurements                                                   & Apache server performance optimization                                & 9/1          & 192  &    \cite{Amiraliminimaldata, menzies2025the, chen2025accuracy, senthilkumar2024can}     \\
1           & Cloud                            & SQL AllMeasurements                                                      & SQL database tuning                                                   & 39/1         & 4,654 &    \cite{Amiraliminimaldata, menzies2025the, senthilkumar2024can}    \\
1           & Cloud                            & X264 AllMeasurements                                                     & Video encoding optimization                                           & 16/1         & 1,153  &   \cite{Amiraliminimaldata, menzies2025the, senthilkumar2024can}   \\
7           & Cloud                            & (rs—sol—wc)*                                                             & misc configuration tasks                                              & 3-6/1      & 196–3,840 &  \cite{Amiraliminimaldata, menzies2025the, senthilkumar2024can, lusstosa2025less, nairMSR18}  \\ \midrule
35          & Software Project Health          & Health-ClosedIssues, -PRs, -Commits                                      & Predict project health and developer activity                         & 5/2-3      & 10,001    &   \cite{Amiraliminimaldata, menzies2025the, senthilkumar2024can, lusstosa2025less,lustosa2024learning} \\ \midrule
3           & Scrum                            & Scrum1k, Scrum10k, Scrum100k                                             & Configurations of the scrum feature model                             & 124/3      & 1,001–100,001 & \cite{Amiraliminimaldata, menzies2025the, lusstosa2025less, lustossa2024isneak} \\ \midrule
8           & Feature Models                   & FFM-*, FM-*                                                              & Optimize number of variables, constraints and Clause/Constraint ratio & 128-1,044/3 & 10,001  &  \cite{Amiraliminimaldata, menzies2025the, lusstosa2025less, lustossa2024isneak}    \\ \midrule
1 &	Software Process Model &	nasa93dem &	Optimize effort, defects, time and LOC	& 24/3 &	93  & \cite{menzies2025the, senthilkumar2024can, lusstosa2025less, lustosa2024learning}\\
1           & Software Process Model           & COC1000                                                                  & Optimize risk, effort, analyst experience, etc                        & 20/5         & 1,001    &  \cite{Amiraliminimaldata, menzies2025the, senthilkumar2024can, lustosa2024learning,chen2018beyond}   \\
4           & Software Process Model           & POM3 (A–D)                                                               & Balancing idle rates, completion rates and cost                       & 9/3          & 501–20,001   & \cite{menzies2025the, senthilkumar2024can, lusstosa2025less, lustosa2024learning, lustossa2024isneak}\\
4           & Software Process Model    & XOMO (Flight, Ground, OSP)                                               & Optimizing risk, effort, defects, and time                            & 27/4         & 10,001    &  \cite{menzies2025the, senthilkumar2024can, lusstosa2025less, lustosa2024learning, lustossa2024isneak,chen2018beyond}  \\ \midrule
3           & Miscellaneous                             & auto93, Car\_price, Wine\_quality                                        & Miscellaneous                                                         & 5-38/2-5   & 205–1,600  &  \cite{Amiraliminimaldata, menzies2025the, senthilkumar2024can, lusstosa2025less, lustosa2024learning} \\ \midrule
4           & Behavioral                       & all\_players, student\_dropout,\newline HR-employeeAttrition, player\_statistics & Analyze and predict behavioral patterns                              & 26-55/1-3  & 82–17,738   &   From  \cite{nyagami_fc25_kaggle_2025, abdullah0a_student_dropout_analysis_prediction_2025, die9origephit_fifa_wc_2022_complete_2025, pavansubhasht_ibm_hr_analytics_attrition_2025}\\ \midrule
4           & Financial                        & BankChurners, home\_data, Loan, \newline Telco-Churn                              & Financial analysis and prediction                                     & 19-77/2-5  & 1,460–20,000 &  From \cite{blastchar_telco_customer_churn_2025, lorenzozoppelletto_financial_risk_for_loan_approval_2025, dansbecker_home_data_for_ml_course_2025, sakshigoyal7_credit_card_customers_2020} \\ \midrule
3           & Human Health Data                & COVID19, Life\_Expectancy, \newline hospital\_Readmissions                        & Health-related analysis and prediction                                & 20-64/1-3  & 2,938–25,000 &   From \cite{dansbecker_hospital_readmissions_2025, kumarajarshi_life_expectancy_who_2025, hendratno_2022}  \\ \midrule
2           & Reinforcement Learning           & A2C\_Acrobot, A2C\_CartPole                                              & Reinforcement learning tasks                                          & 9-11/3-4   & 224–318  &     \\ \midrule
5           & Sales                            & accessories, dress-up, Marketing\_Analytics, socks, wallpaper            & Sales analysis and prediction                                         & 14-31/1-8  & 247–2,206 &   From \cite{jessicali9530_animal_crossing_new_horizons_nookplaza_dataset_2021, jackdaoud_marketing_data_2022, syedfaizanalii_car_price_dataset_cleaned_2025}  \\ \midrule
2	& Software testing	& test120, test600	& Optimize the class	& 9/1	& 5,161\\ \midrule

127         & \textbf{Total}                            &                                                                          &                                                                       &              &        &      
\end{tabular}
\end{adjustbox}
\endgroup
\end{table*}
One lesson from this paper's analysis of Table~\ref{combinedtable} is that there are many examples where AI for SE  can be dramatically simplified.  
 Human cognitive biases (see \S\ref{cog}) mean we routinely  ignore simpler solutions.  Hence we fear there
 are many other cases where research  has recommended needlessly    complex solutions.
 We hope our limited results to date   encourage others   to take up a new data-light challenge: 
\vspace{0.2cm}
 
 \begin{quote}
\ {\bf The Data-Light Challenge}: {\em When will a handful of labels suffice for SE tasks?} 
\vspace{0.2cm}
 \end{quote}
 
\color{black}
The rest of this paper is structured as follows:
\begin{enumerate}
\item
We establish {\bf why simplicity matters}. Beyond obvious computational benefits, cognitive and pragmatic factors make conclusions from minimal data  valuable for real-world SE practice.
\item
We formalize the {\bf mathematics of simplicity}; i.e. the number
of samples needed for different  goals.  Some SE tasks truly need thousands to millions of samples (or more). But other tasks are much simpler such as {\em Near Enough Optimization}. In NEO,
(a)~near enough is good enough, and (b) supposedly better solutions are indistinguishable from the
best seen so far.  For example, in project management, there is much variability in the effects of different decisions. Hence, many micro improvements can be ineffective. Software manager seeks large ``big picture'' changes with large impact. In  NEO problems, near-ties are good enough, and search halts once no
option has a clear additional benefit.    We show mathematically that NEO 
can terminate after   $\approx 10^1$ samples.
\item  We demonstrate the {\bf prevalence of NEO}.   Recently.
Chen and Menzies have been collaborating to curate the MOOT repository (Many Multi-Objective Optimization Tasks\footnote{ref:~\cite{menzies2025moot}; \url{http://github.com/timm/moot/optimize}}), which contains datasets from recent SE optimization papers.
After two releases, MOOT contains the 127    NEO-compatible tasks of Table~\ref{combinedtable} .   Since these  tasks come from recent literature, we assert that NEO is a prevalent and current SE challenge.
\item 
\color{black}
\here{A4}\BLUE We present {\bf experiments with NEO} across multiple state-of-the-art optimizers on MOOT data. Using only a few dozen labeled examples, simple approaches—diversity sampling, minimal Bayesian learners, their non-parametric distance-based variant, even random probes—achieve over 90\% of  optimal performance. These lightweight strategies perform as well as  advanced optimizers like Sequential Model-based Algorithm Configuration (SMAC) \cite{hutter11smac}, Tree-structured Parzen Estimator (TPE) \cite{bergstra11TPE}, and Distributed Evolutionary Hyperband (DEHB)~\cite{awadijcai2021p296}. While certain applications such as mission-critical tasks demand higher precision, these outcomes observed after just dozens of trials would suffice for the engineering tasks considered in this study. Moreover, given the label shortages discussed in \S\ref{datlabel}, these results may represent the practical ceiling achievable without impractical labeling efforts.\color{black}
\item
Finally, we offer a test for  {\bf when simplicity  might fail}. This test uses a previously unreported {\em BINGO effect}: when SE data splits into $n$ buckets across $d$ dimensions divided into $b$ bins, most data concentrates in surprisingly few buckets (i.e.  $n \ll b^d$). In one study with 10,000 rows, we expected $b^d = 4096$ buckets from $d = 4$ dimensions split into $b = 8$ bins, but found only 100 occupied buckets. When data exhibits this BINGO effect, evaluating millions of options becomes unnecessary (since only 100 distinct scenarios exist). We recommend testing for BINGO on new datasets and avoiding our simple methods when data spreads across too many buckets.
\end{enumerate}
In summary, the contributions of this paper are:
\begin{itemize}
\item A new challenge for the SE research community: the {\em data-light challenge}, asking when and under what conditions a small number of labels suffices.
\item Recognition of a counter-intuitive anomaly: in the era of big data, we find  some cases where minimal data yields strong results.
\item We document mathematical results identifying when few labels will (or will not) suffice.
\item A library of simple baseline algorithms for optimization under severely limited labeled data.
\item Empirical results showing that optimization with a few labels   can work as well as the  prior state-of-the-art. 
\item A  method for predicting when {\em not}
to use simple methods (see the BINGO effect of \S\ref{bingo_eff}).
\item All materials released under open-source licenses at \url{https://github.com/KKGanguly/NEO}. 
\end{itemize}

\subsection{Frequently asked questions}
Before we begin, we digress to discuss  some  frequently asked questions about this work.

\BLUE
 \here{A5}
{\bf FAQ1: Does these results show that all AI tasks can be simplified?}.
No. This study shows that  {\em some} AI for SE problems, such as the ones studied here, have remarkably simple solutions. We do not claim that {\em most} AI for SE can be simplified using these methods.  That said, we hope this study motivates others to explore simpler methods in their research. \color{black}

{\bf FAQ2:  What do we mean by ``optimal''?} 
Here, ``optimal'' means ``reference optimal,'' (which is a term from empirical algorithms 
literature~\cite{McGeoch2012,cohen1995empirical}). 
The reference optimal refers to the 
best solution observed so far.  The reference optimal may not be the 
true optimal. However,    in many engineering cases, the true
optimal may be unknown.

{\bf FAQ3: What is novel about this work? Is it not widely accepted that simple 
software solutions can  sometimes suffice?} Perhaps not. If the SE community widely 
believed this, our literature would show complex methods routinely 
compared to simpler ones. It does not, especially for AI in SE. For 
instance, in a recent survey of 229 SE papers on LLMs, only 5\% compared to 
simpler approaches~\cite{Hou24}.

Another novelty in this paper is that  the labeling reductions reported here are   far larger than seen elsewhere. 
For example, in our billing10K data 
set (10K rows, each describing 88 design choices for an online  billing system), our algorithms found 
options within 80\% of optimal using just 50 labels (i.e. just 0.5\% of the data). 
By contrast,  when working on software vulnerability detection, Zhu et al. needed 30\% labeled rows   and
Zhang et al.~\cite{nsglp} required labels on  10–30\% for their defect prediction data. 

Another simplicity measure is CPU cost. One task explored in this paper  is 
hyperparameter optimization (HPO), i.e. automatically finding control 
parameters for the data miners used in software analytics. HPO is usually slow. For example, a study by
Eggensperger et al.~\cite{eggensperger2013},  random search, 
Bayes, Hyperband, etc. needed labeling budgets of 4K,12K, and 20K. Collectively, these runs consumed 22.5 CPU 
years. Our algorithms, on the other hand,  are  faster since they explore far less data.  For example: (a)~the above analysis of billing10k took 0.5 seconds (on a standard desktop Mac without using any GPU); (b)~our algorithms run more than  two orders of magnitude faster than standard state-of-the-art methods (see the runtime results of Figure~\ref{fig:perf}).

{\bf FAQ4 :  Why seek a new simpler solution when some current
solution, albeit complex, is available?}
Out next section argues that searching for   simpler systems has important practical and cognitive
implications.

\section{Motivation: Why Simplicity Matters}\label{why}

Many authors have commented that simplicity offers  significant benefits
including less compute usage, reduced energy cost~\cite{calrero15}, decreased pollution from energy production~\cite{monserrate2022cloud}, easier explanation~\cite{veerappa2011understanding}, faster tuning~\cite{fu2016tuning}, simpler customization, reproducibility, and more trustable results~\cite{rudin2019stop}. 

For millennia,
simplicity has been advocated  by philosophers and researchers. For example, in  130 A.D., Ptolemy wrote {\em ``We consider it a good principle to explain
the phenomena by the simplest hypothesis possible''}.
A thousand years later, William of Occam made  a similar point ({\em ``entities must
not be multiplied beyond necessity''}). Many researchers  show how complex systems might   reduce to simpler parts:
\begin{itemize}
  \item {\bf 1902}, {\em PCA}: reduce  data to a few principal components \cite{pca}.
  \item {\bf 1960s}, {\em Narrows}: guide search via a few key variables \cite{Amarel1986}.
  \item {\bf 1974}, {\em Prototypes}: speed up $k$-NN by reduce rows to a few    exemplars \cite{chang74}.
  \item {\bf 1984}, {\em JL lemma}: random projection to $k=O(\varepsilon^{-2}\log n)$ dimensions can preserve pairwise distances to within some error $(1\pm\varepsilon)$ \cite{johnson1984extensions}.
  \item {\bf 1986}, {\em ATMS}: only focus diagnosis on  core assumptions \cite{atms}.
  \item {\bf 1994}, {\em ISAMP}: a few restarts can explore large problems spaces \cite{Crawford:1994}.
  \item {\bf 1996}, {\em Sparse coding}: learn efficient, sparse representations from data which inspired dictionary learning and sparse autoencoders \cite{olshausen2004sparse}.
  \item {\bf 1997}, {\em Feature selection}: ignore up to 80\% of features \cite{kohavi97}.
  \item {\bf 2002}, {\em Backdoors}: if we first set  a few variables, that cuts exponential tine to polynomial \cite{backdoor}.
  \item {\bf 2005}, {\em Semi-supervised learning}: data can be appoximated on a much lower-dimensional manifold \cite{zhu05}.
  \item {\bf 2009}, {\em Active learning}: only use most informative rows  \cite{settles2009active}.
  \item {\bf 2003–2021}, {\em SE “keys”}: a few parameters govern many SE models \cite{me03a,me07a,me21a,menzies2008implications}.
  \item {\bf 2010+}, {\em Surrogates}:  first, build small models to  label the rest of the data \cite{zul13,guo13}.
  \item {\bf 2020s}, {\em Distillation}: compress large LLM models with little performance loss \cite{shi22,yang24}.
\end{itemize}
Strangely, despite all the above,
simplification studies are not   common. As mentioned above,  in a recent survey of 229 SE papers that explored LLMs for SE, 
 only 5\% of papers compared their results against a simpler approach~\cite{Hou24}. 

\subsection{Cognitive Issues}\label{cog}
\begin{wrapfigure}{r}{1.5in}
\begin{center}
\includegraphics[width=1.5in]{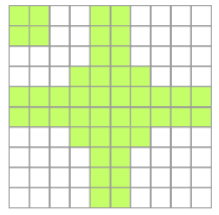}

\includegraphics[width=1.5in]{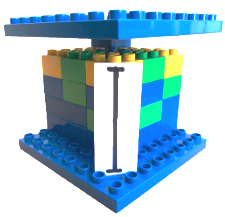}
\end{center}
\caption{Examples where simple subtractive changes can improve a design. From~\cite{adams2021people,Fillon25}.}
\label{change}
\end{wrapfigure}

We conjecture that  this  lack of simplicity studies reflects a widespread
cognitive bias.  Adams and  Fillon et al.~\cite{adams2021people,Fillon25},
 studied problems like Figure~\ref{change} where
 subjects were asked 
to (e.g.) make a
green logo more symmetrical or make a LEGO tower hold   more weight. 
They found that~ {\em people systematically overlook subtractive changes}.
Subjects  where  asked to (a)~make
the green logo (at top) more symmetrical; or (b)~redesign the LEGO tower (at bottom)  to support more weight. Those subjects
rarely
removed the top-left green tiles, or the tiny collar that precariously
supports the upper LEGO tray.
Fillon et al.~~\cite{Fillon25} replicated that work
and found the ratio of additive to subtractive changes was 1155 to 297 (about 4:1). To say that another way,  {\em humans  needlessly complicate their designs}.

In the recent SE literature, there is evidence that this cognitive bias means that researchers
often overlook simpler solutions, especially when it comes to LLMs. For example,
Tawosi et al. report that 
when predicting effort within agile projects, traditional text mining methods (SVM and TF\*IDF) 
resulted in better predictions that a prior study that used deep learners.
Further, those traditional text mining methods ran over 100 times faster than the deep
learners~\cite{Tawosi23},
Similarly, Majumder et al. showed that
simply pre-clustering the data with K-Means  achieves comparable accuracy to deep learning for software defect prediction while running 500+ times faster~\cite{majumder2018500+}.
In other work  on
data synthesis, Ling et al. showed that  parametric statistical methods performed as well (and run much faster)
than GAN-based methods~\cite{Ling24};
Further, Fu et al. have shown that 
very fast tuning works surprisingly well for hyperparameter optimization for   multiple software engineering prediction tasks~\cite{Fu17}.
Furthermore, despite advances in deep tabular learning,  for tabular data, traditional ML methods often achieve comparable or superior performance with significantly less computational overhead~\cite{somvanshi2024survey}. 
Tree-based models (Random Forest, XGBoost) consistently outperform deep neural networks on many tabular datasets, especially those with more than 10,000  samples ~\cite{grinsztajn2022why}.


\subsection{Data Labeling Issues}\label{datlabel}
Apart from cognitive concerns,  exploring less data is also of 
practical benefit for addressing the {\em labeling problem}.
 Consider an agent exploring the world:
 \begin{itemize}
 \item {\em Choices} may be easy to spot;
 \item But {\em consequences} are slow to measure.
 \end{itemize}
 For example, in a used car lot, 
a shopper can scan hundreds of used cars in minutes.
However, it would require hours of driving to each car's gauge fuel economy.  

Formally, this is the \emph{labeling} problem. Suppose our goal is to learn how choices $x$ affect consequences $y$. To learn the function $y = f(x)$, we need $(x,y)$ pairs; i.e.  pairs of 
choices, labeled with their consequences.   

\begin{table}[!t]
\caption{For $ y=f(x)$, it is often   cheaper to  collect $x$ values     than the 
associated $y$ values.}

{\footnotesize 
    \begin{tabular}{p{2.6in}|p{2.6in}}
   Find $X$ &  Find associated $Y$ values\\
 \rowcolor{black}    \textcolor{white}{   It can be very  quick to ...} & \textcolor{white}{ It is a much    slower task  to... }\\ 
   
Mine GitHub to find  all the  distributions of code size, number of dependencies per function, etc.
& Discover (a)how much that software  could be sold on the market or (b)  what is the time required
to build this kind of software\\

\rowcolor{gray!20} Count the number of classes in a system. &
Negotiate with an organization permission to find how much human effort was required to build and maintain that code.\\
       List   design options; e.g. 20 binary choices is $2^{20} > 1,000,000$  options.  &  Check all those options with a group of human stakeholders.\\
       
 \rowcolor{gray!20}       List the   configuration parameters for some piece of software. & Generate a separate executable for
       each one of those parameter settings, then run those executable through some test suite.\\    
       
      List the   controls of a data miners used in software analytics  (e.g. how many neighbors to use in a k-th nearest neighbor classifier). & Run a grid search looking for 
the best settings for some local data.\\

 \rowcolor{gray!20} Generate test case inputs (e.g.) using some grammar-based fuzzing.& 
Run all all those tests. It can be even slower for a human to check through all those results
looking for  anomalous behavior.\\
 
    \end{tabular}}
    \label{slow}
\end{table}

An interesting feature of the values of $x$ and $y$ is that $x$ is much cheaper to collect that $y$.
Table~\ref{slow} shows numerous examples in software engineering where a data scientist would have ready access to $x$ values, but far less access to $y$ values.
The problem with $y$ values is that while there are many ways to find those values,  each have their  limitations:

 \begin{enumerate}
 \item
 In {\bf expert labeling},   subject matter experts offer $y$ labels. Label quality degrades when experts are rushed to process large corpora\cite{easterby1980design}. More careful interviews  to label, say, 10 examples with ten attributes can require one to two hours per session\cite{KingtonAlison2009, lustosa2024learning}, repeated two or three times per week~\cite{valerdi2010heuristics}\footnote{If this seems too slow, then we remind
that reader that this is consistent with one of the heuristics for manual knowledge acquisition from the 1980s; i.e. `5-10 expert rules per day''~\cite{menzies1992expert}.}.
\item
 {\bf Historical logs} provides ``free'' labels, but often includes errors. For instance, Yu et al.\cite{yu2020identifying} found 90\% of technical debt entries marked ``false positive'' were actually correct. Similar problems exist in security datasets\cite{wu2021data}, static analysis tools\cite{kang2022detecting}, and defect datasets\cite{shepperd2013data}.
\item
{\bf Automated labeling}  models can generate $y$ labels, but simplistic labeling models    can be misleading  (e.g. simple regex-based labeling methods can rely on dubious   keyword heuristics~\cite{kamei2012large}).
Recent studies on automatic labeling at MSR'25 report that   LLMs, while promising, remain assistive not authoritative labeling agents\cite{ahmed2025can}.
Even after decades of use, some automated labeling models remain problematic;
e.g. the  SZZ algorithm is the de facto standard for labeling bug fixing commits and finding inducing changes for defect prediction data~\cite{Sliwerski05}. But there are many issues with the labels it generates
and some researchers are negative 
about SZZ ever becoming a truly authoritative source~\cite{herbold2022problems}. 
The main point here is that 
even seemingly sophisticated automated labeling methods have to be   certified--  which means   automated labeling  is dependent on experts or logs for their evaluation. 
\end{enumerate}
\subsection{Implications for this Paper}\label{hardlabel}
Since using historical logs and automated labeling is   problematic, we seek to better support expert
 labeling. Experts are busy people so we must not bother  with excessive questions.
 To that end, the experiments of this  paper asks ``just how few labels are required to build a model''.   
 
 To support domains like in Table~\ref{combinedtable}, we seek solutions
 relying mostly on $x$ values, and very little on $y$ values.  
 Also, consistent with point \#1 (above), we ask  what can be done with 100 labels or less.


\section{The Mathematics of Simplicity}\label{maths}

To summarize the previous section, it would be useful if software engineering   problems could be solved with very little data. But is there any evidence for that hope, or is it just wishful thinking?

This section answers this question by documenting three mathematical results that discuss sample
requirements for different problem types, from
least to most ambitious.

The least ambitious goal is {\bf Near Enough Optimization} (NEO); i.e.
seek solutions that are ``close enough'' to the 
optimal. In NEO, (a) near enough is good enough, and (b) supposedly
better solutions are indistinguishable from the best seen so far.

To model NEO, we assume  exploring a table of data   where, for each row,
the $x$ values are known
and $y$ values are hidden. We randomly sample rows,
assuming: (i) solutions are randomly shuffled ensuring even spacing,
and (ii) there exists some small $\varepsilon$ below which the $y$ values
of two rows  
are indistinguishable. The second assumption follows Deb et al.'s
epsilon domination criteria~\cite{deb2005epsilon}, where $\varepsilon$
divides the problem space into hypercubes requiring only one sample
per cube.

A geometric distribution model represents the waiting time until first
success as independent Bernoulli trials~\cite{ross2014introduction}.
For a process where the probability of finding a solution within
$\varepsilon$ of optimal is $\varepsilon$, the expected trials until
first success is $1/\varepsilon$. To achieve confidence $C$ of
observing at least one success within $n$ trials then 
C=$1-(1-\varepsilon)^n$ (assuming Bernoulli trials)\footnote{For Bernoulli 
trials where the probability of finding a solution within $\varepsilon$
of optimal is $\varepsilon$, the probability of failure for $n$ trials
is $(1-\varepsilon)^n$. Setting $1-(1-\varepsilon)^n \geq C$ and
solving gives $(1-\varepsilon)^n = 1-C \Rightarrow n \log(1-\varepsilon)
= \log(1-C) \Rightarrow n = \tfrac{\log(1-C)}{\log(1-\varepsilon)}$.}. Solving for $n$ 
leads to 
\begin{equation}\label{eqneo}
n_\mathit{neo} > \tfrac{\log(1-C)}{\log(1-\varepsilon)}
\end{equation}
 Near Enough reasoning stops after finding one satisfactory
solution. An alternative is finding the optimal solution among all
known candidates. This is the {\bf Best Arm} problem in reinforcement
learning~\cite{lattimore2020bandit}.
For such Best Arm problems,  the
Hoeffding's inequality provides theoretical foundations for sample
complexity bounds~\cite{hoeffding1963probability}. To distinguish
between $A$ alternatives with accuracy $\varepsilon$ and confidence
$C$, we require approximately $(2/\varepsilon^2)\log(2A/(1-C))$
samples per alternative. If $A$ represents candidate actions and
$\varepsilon$ is the accuracy tolerance, it follows\footnote{By
Hoeffding's inequality, for bounded random variables the probability
that the sample mean $\hat{\mu}$ deviates from the true mean $\mu$ by
more than $\varepsilon/2$ is $P(|\hat{\mu} - \mu| > \varepsilon/2)
\leq 2\exp(-2n\varepsilon^2/4) = 2\exp(-n\varepsilon^2/2)$, where $n$
is the number of samples. For confidence $C$ per arm (using union
bound over $A$ arms): $2\exp(-n\varepsilon^2/2) \leq (1-C)/A$.
Solving for $n$: $\exp(-n\varepsilon^2/2) \leq (1-C)/(2A)
\Rightarrow -n\varepsilon^2/2 \leq \log((1-C)/(2A)) \Rightarrow n
\geq \frac{2}{\varepsilon^2}\log\frac{2A}{1-C}$.} that:

\begin{equation}\label{eqbest}
n_{\mathit{bestarm}} \geq  \frac{2}{\varepsilon^2}\log\frac{2A}{1-C}
\end{equation}

Both Near Enough and Best Arm assume known reward functions.
But what if the reward function is unknown? Consider autonomous
vehicles: how many samples can certify software against all possible
future legal requirements for driveless cars? This is {\bf Reward-Free} reinforcement
learning problem. A \textit{reward function} assigns numerical scores to
outcomes. Future regulations might require optimizing for undefined
criteria—speed, efficiency, comfort, or safety compliance not yet
specified.  Reward-Free reasoning requires
extensive state-space exploration across all potential future
objectives.

Deriving  the Reward-Free sampling complexity is complex~\cite{jin2020reward}
so we just summarize that result.
For $S$  discrete states, $A$  candidate actions,
and a planning horizon of  $H$:

\begin{equation}\label{eqfree}
n_{\mathit{reward\; free}} \geq S^2AH^3\log(1/(1-C))/\varepsilon^2
\end{equation}

Using Equations~\ref{eqneo}, \ref{eqbest}, and \ref{eqfree},
we can see that 
different problems lead to vastly different sampling requirements.
For example, assuming:
\bi
\item $H=1$; i.e. one-step plans.
\item $\varepsilon=0.05$; i.e. we are content to reach 95\% of the  optimum.
\item $A=(b^d)$ where $b,d=6,6$ i.e. ~PCA or  random projections has reduced
our tabular data
to $d=6$ dimensions divided into $b=6$ bins. Assuming only $d=6$ is a somewhat
optimistic assumption but, as we shall see, even with this optimism, Reward-free reasoning will be seen to be our hardest problem. 
\ei
Under these assumptions, we see that
  two   problems are extremely data hungry, but NEO conforms to  
constraints seen in \S\ref{hardlabel}:
\begin{itemize}
\item Equations~\ref{eqneo} says  Near Enough optimization needs at least 60 samples;
\item Equation~\ref{eqbest} says Best Arm identification needs at least  6,992 samples;
\item Equation~\ref{eqfree} says Reward-Free reinforcement learning needs at least $4.6 \times 10^{15}$
samples.
\end{itemize}
In summary:
\vspace{-0.3cm}
\begin{conclusion}{Mathematically:}
an argument can be made that  
NEO-compatible software engineering problems can be solved with just
a few dozen samples.
\end{conclusion}


\begin{figure}[!t]
\centering

\begin{minipage}{3in}
\centering
Figure~\ref{badchoices}a: software accumulates config.
\end{minipage}
\hspace{0.15in}
\begin{minipage}{2in}
\centering
Figure~\ref{badchoices}b: much config is ignored
\end{minipage}
\vspace{0.5em}

\begin{minipage}{3in}
\centering
\includegraphics[width=3in]{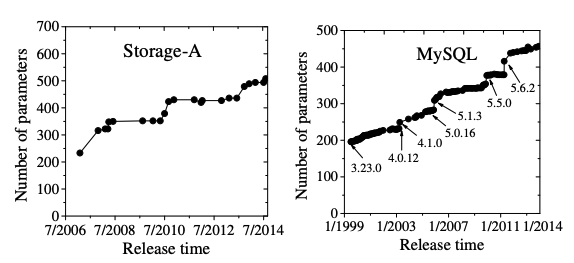}
\end{minipage}
\hspace{0.15in}
\begin{minipage}{2in}
\centering
\includegraphics[width=2in]{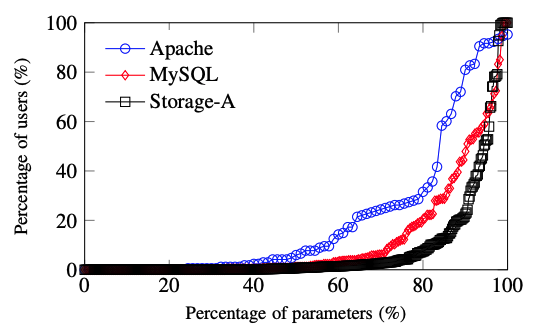}
\end{minipage}
\vspace{-0.3cm}
\caption{
The number of configuration parameters over time (left)
but few of  those parameters are actually used  by people (right). From~\cite{Xu2015}. }
\label{badchoices}
\vspace{-0.3cm}
\end{figure}
\section{Prevalence of NEO}
\here{A6}\BLUE This section argues that there are several NEO-compatible problems in SE, occurring across a range of tasks including configuration tuning, cloud optimization, project and process modeling, and numerous empirical software analytics domains.

Modern software engineering offers a vast array of choices relating to software systems.  \color{black}Every software
layer (languages, libraries, frameworks, deployment pipelines) can be characterized by a plethora of   configuration files containing an overwhelming number of flags and
defaults.  It is not a rhetorical flourish to say that our software
has more options than stars in the sky.
One
  open-source database we studied (SQLite) had 460 binary options.
  $2^{460}$ is larger than the $2^{80}$ stars in the observable universe~\cite{doe2023personal}.  That number of configuration options dwarfs Avogadro’s constant and makes
brute-force search   impossible.


In Figure~\ref{badchoices}, Xu et al.~\cite{Xu2015} document the administrative pain of adaptive
reconfiguration in cloud systems: even modest workloads required juggling
throughput, latency, and energy budgets simultaneously. Developers face
dilemmas that cannot be solved by ``common sense'': more code at less cost,
faster queries with lower wattage, higher reliability without expanding test
budgets. Apel~\cite{Apel2020} and Zhou~\cite{Zhou2011} catalog dozens of
similar contradictions across software product lines and distributed systems.   
Based on that experience, they warn:

\centerline{\em Configurability is a \underline{liability} without tool
support.}


\here{A7}\BLUE
Beyond configuration, SE contains several other NEO-compatible tasks. Recent SE research reveals that multi-objective optimization problems in SE extend well beyond software configuration \cite{menzies2025moot}. To support research in this broader space, 
researchers   Chen and   Menzies
 have been collecting data sets in their MOOT repository.  
 MOOT covers wide array of tasks such as - 
 \begin{wrapfigure}[30]{r}{2.5in}
{\scriptsize
\begin{alltt}
\vspace{0.2cm}
 (a) Configuration tuning:
  x = independent values     | y = dependent values
-----------------------------|-----------------------
  Spout_wait, Spliters, Cntr. | Throughput+, Latency-
     10,        6,       17  |    23075,    158.68
      8,        6,       17  |    22887,    172.74
      9,        6,       17  |    22799,    156.83 
    [Skipped],  ...,      ...|          ...,    ...
  10000,        1,       10  |   460.81,    8761.6
  10000,        1,       18  |   402.53,    8797.5 
  10000,        1,        1  |   310.06,    9421
-------------------------------------------------------

(b) Defect prediction 

x = independent values    | y = dependent values
--------------------------|---------------------
Numobs, Length, ..., Churn| Class+
 3317,   2990,  ...,  0.2 |   1
 5873,   1923,  ...,  0.4 |   2
 2357,   1333,  ...,  0.0 |   2
 [Skipped],     ...,  ... |   ...
 1104,    891,  ...,  0.4 |   3
  204,     38,  ...,  0.4 |   3
 1939,   1236,  ...,  0.1 |   3
-------------------------------------------------------

(c) Hyperparameter Tuning for Software Health Prediction

x = independent values         | y = dependent values
------------------------------|----------------------
N_est, Min_leaf, ..., Min_imp.| MRE-, ACC+, PRED40+
   40,     1,    ...,    0.25 | 0.205, 0.826, 1.00
  120,     1,    ...,    4.75 | 0.191, 0.756, 1.00
   10,     1,    ...,    6.75 | 0.000, 0.790, 0.86
 [Skipped],      ...,    ...  |   ...,   ...,  ...
   10,    11,    ...,    5.25 | 0.925, 0.391, 0.00
   90,    12,    ...,    5.00 | 0.911, 0.397, 0.00
  130,     9,    ...,    9.75 | 0.914, 0.395, 0.00
\end{alltt}
}
\caption{Example MOOT datasets.}\label{moot}
\end{wrapfigure} 
\bi
    \item \textbf{Software System Optimization:}
  Tuning runtime parameters to optimize metrics such as throughput, latency, and energy (Specific software configuration, PromiseTune \cite{chen2026promisetune} datasets, cloud).

  \item \textbf{Product Line Engineering:}
  Solving high-dimensional constraint satisfaction problems for product line feature models (Feature Models, Scrum).

  \item \textbf{Software Project Health Forecasting:}
  Optimizing predictive models for pull request closures, issue resolution, and developer commits (Software Project Health).

  \item \textbf{Defect Prediction \& Mitigation:}
  Optimizing process simulations to minimize defect density (nasa93dem, XOMO in Software Process Model)
  and test feature selection to predict bugs (Software Testing).

  \item \textbf{Software Process \& Cost Estimation:}
  Simulating agile and waterfall processes to balance risk, effort, and cost (POM3, COC1000).

  \item \textbf{Cross-Domain Generalization Benchmarks:}
  A diverse control group of non-SE optimization tasks (Financial, Medical,
  Behavioral, Reinforcement Learning etc.).
\ei

 Fig.~\ref{moot}
shows the    structure  of  MOOT's 127 tasks.
All tables have $x$ inputs and $y$ goals where
(a) the first row shows the column names;
(b) numeric columns start with uppercase, all others are symbolic;
(c)~goal columns (e.g. Fig.~\ref{moot}.a {\em
Throughput+, Latency-}) use +/- to show  maximize and minimize.
Goals vary by task.
In  Fig.~\ref{moot}.a configuration parameters such as {\em Spout\_wait, Spliters, and Counters} are tuned to optimize Throughput and Latency,
In Fig.~\ref{moot}.b software metric sets are identified to best predict defect classes.
In Fig.~\ref{moot}.c, machine-learning hyperparameters are adjusted to improve prediction accuracy and reduce error in software health prediction.




Tables like Fig.~\ref{moot} are compatible with the data model used to explore Near Enough optimization in \S\ref{maths}.
 MOOT’s first release contained 39 datasets, and the latest release expands this to 127 datasets drawn from recent SE literature. This abundance of   problems  makes us assert: \color{black}
 \FloatBarrier
 \begin{conclusion}{Observation:}
 NEO-compatible problems are prevalent in the SE literature.
 \end{conclusion}
 
 \here{A8}\BLUE We hasten to add that we do not claim that all SE problems are NEO-compatible. For tasks
 that are not NEO-compatible, see Best arm and Reward-free optimization in
 \S\ref{maths}. \color{black}

\section{Experiments with NEO}\label{expr}

Using data from   MOOT,
this section compares simple and complex methods for Near Enough optimization.
It will be shown that simple optimization methods based on a few dozen labels   work as well, and sometimes better, than
  state-of-the-art algorithms  (that are much more complex). 

 \subsection{Measuring Successful Optimization}\label{d2his}
NEO reasons over  rows of data containing $x$ input independent choices and $y$   dependent goals.   When $|y|>1$, NEO becomes  
{\em multi-objective problem}~\cite{zapotecas2020multi} that must struggle to maximize some goals
while minimizing others (e.g. build 
{\em better} software, {\em faster} and 
{\em cheaper}).  

 To score the results of our multi-objective optimization, we use {\em distance to 
 heaven} where ``heaven''  is the best value of each   $y$-values. 
E.g. in Figure~\ref{moot}, the $y$ columns are labelled 
{\em Throughput+, Latency-} which means we want to maximize the first and minimize the second.

After normalizing the dependent values to a 0..1 range (min to max), then {\em heaven} -is $\{1,0\}$. Any row in that data set can be scored on how far away it falls from that point.  Note that, for this measure,
{\em lower} values are {\em better},
  since {\em smaller} values are {\em closer} to the ideal.

For every dataset in Table~\ref{combinedtable},
there exists a mean $\mu$ and a {\em min} distance to heaven. After each run, our  optimizers return a row with a distance to heaven score of $x$. After 20 repeated runs,
these optimizers return best rows
with a mean score of $x$. 
We define:
\vskip 0.1cm
\begin{equation}\label{delta}
\Delta = 100 * \left(1 - \frac{x-\mathit{min}}{\mu - \mathit{min}} \right)
\end{equation}
\vskip 0.1cm
For our methods,   $\Delta \le 100$ and
\bi
\item
$\Delta = 0$ is a failure since  optimization does not improve  initial conditions;
\item
  $\Delta=100$ is a    success since  optimizers  find the best
row with {\em min} distance to heaven.
\ei
For the reader familiar with the optimization literature, we note that there are many other ways to measure {\em benefit} in multi-objective reasoning.
A recent IEEE TSE article by Chen et al. \cite{Li22} reviewed various multi-objective optimization performance measures like Hypervolume, Spread, Generational Distance, and Inverted Generational Distance. We consulted with
those authors on this paper and they offered the opinion~\cite{doe2023personal} that their measures are inappropriate
this paper
due to our focus on minimal labeling.
For instance, our approach would inherently result in low hypervolume since we generate very few solutions.

\subsection{Algorithms}
Optimization strategies can differ between {\em pool-based} and 
{\em membership query} inference~\cite{settles2012active}. 
Pool-based methods search existing datasets, while membership query 
methods interpolate between rows to propose new solutions. In the following sections, we describe the state-of-the-art algorithms in these two categories. Detailed pseudocodes for these algorithms are available online  at \url{https://github.com/KKGanguly/NEO/blob/main/supplementary.pdf}.

\subsubsection{Membership Query Inference}
These algorithms construct arbitrary rows (not from pre-defined pools) and 
request their labels. \here{A9}\BLUE Examples include evolutionary programs 
that mutate/combine solutions, and many model-based optimizers that synthesize new candidates using surrogate models or probabilistic acquisition functions. We selected four state-of-the-art representative membership query inference algorithms which are  \underline{\em SMAC}, \underline{\em TPE},    and \underline{\em DEHB}.

\underline{\em SMAC} was selected as an representative for Random Forest-based BO. Unlike standard Gaussian Processes, SMAC utilizes Random Forests to model the objective function. This enables supporting categorical parameters and non-continuous landscapes. To propose a new solution, SMAC maximizes the Expected Improvement (EI) acquisition function. Since Random Forests are not differentiable, SMAC performs this maximization by generating a large set of random samples and performing local searches (mutations) around the best points found so far to propose the next configuration. In recent experiments \cite{lindauer2022smac3}, SMAC has been shown to outperform BOHB~\cite{awadijcai2021p296}, Hyperband \cite{li2018hyperband}, ParamILS \cite{hutter2009paramils}, and Dragonfly \cite{kandasamy2020tuning}.  

\underline{\em TPE} was selected as a representative of tuners widely used in industry (e.g. Hyperopt and Optuna). It uses  density optimization Bayesian optimization. It inverts standard Bayesian inference. Instead of modeling $p(y|x)$, it models $p(x|y)$ using two kernel density estimates, $l(x)$ for "best" samples and $g(x)$ for the "rest". The generation of new candidates is done by maximizing the ratio $l(x)/g(x)$. Thus, it effectively interpolates to find points highly likely to be "good". TPE outperforms GP, and recent studies find that it performs competitively with BOHB and SMAC on several low-dimensional and continuous benchmarks \cite{bergstra11TPE, awadijcai2021p296}.


\underline{\em DEHB}, a representative for evolutionary multi-fidelity optimization, combines differential evolution search with Hyperband's adaptive 
resource allocation \cite{awadijcai2021p296}.
 Hyperband \cite{li2018hyperband} uses {\em successive 
halving}~\cite{jamieson2016nonstochastic} across varying budgets, 
running configurations with small budgets, discarding the worst 
performers, promoting remainder with larger budgets.
 Differential Evolution \cite{storn1997differential} 
generates solutions via mutation/crossover on population vectors, 
selecting superior vectors. Unlike bit-by-bit genetic algorithms, 
DE extrapolates between known good solutions.
  DE candidates are evaluated by Hyperband using successive 
halving to allocate resources and prune poor configurations.
DEHB is the current 
state-of-the-art combining evolutionary methods (DE), 
multi-fidelity methods (Hyperband), and iterated racing. It outperforms BOHB~\cite{awadijcai2021p296}, which outperforms 
Hyperopt~\cite{bergstra2015hyperopt}. 


\subsubsection{Pool-based Inference}
Pool-based inference assumes large unlabeled pools, selecting the most 
informative examples for labeling. Examples include 
\underline{\em RANDOM} and the  \underline{\em LINE} and 
\underline{\em LITE}  methods described below.
(Aside: We claim no special cleverness for RANDOM/LITE/LINE. We only use them to  demonstrate
that simple algorithms work remarkably well with few labels.)

Given budget $B$, \underline{\em RANDOM} labels any $B$ rows, 
sorts by distance to heaven,  and returns best.

\here{A10}\BLUE
\underline{\em LINE} samples points different from those chosen previously. 
Diversity sampling explores data well~\cite{ijcai2024p540}, often 
cluster-based \cite{yehuda2018cluster}. LINE uses  the K-Means++ 
centroid finder~\cite{Arthur2007kmeansplusplus}: 
\bi
\item Given already selected rows $S$, a new row $x$ is chosen with probability proportional to, 
\vskip 0.1cm
\begin{equation}
    D(x, S)^2 = \left( \min_{s \in S} \| x - s \| \right)^2
\end{equation}
\vskip 0.1cm
\item This produces a set of points covering high variance regions of the dataset.
\item Diversity sampling improves search space coverage and reduces redundant label acquisition.
\ei

\underline{\em LITE} was inspired by Bergstra's Tree of Parzen 
Estimators~\cite{bergstra2012random}. It was an experiment in simplifying TPE and adapting it to fixed pools. LITE's mechanism: 
\bi 
\item It begins by labeling a small number of rows.
\item Instead of Parzen estimators, it sorts and divides labeled data to $\sqrt{N}$ {\em good} and $N-\sqrt{N}$ {\em rest}. 
\item It then builds a two-class classifier (e.g. Naive Bayes) reporting the likelihood of unlabeled examples 
being good $g$ or rest $r$. 
\item LITE labels the  next row that maximizes the acquisition function $g/r$. Updated labeled rows are 
sorted and the process repeats. 
\item If the {\em good} set grows beyond $\sqrt{N}$, the lowest-ranked items are moved to the {\em rest} set.
\ei
Other acquisition functions for LITE were explored but they showed 
statistically indistinguishable behavior in our results.

\underline{\em EZR} is a variant of \underline{\em LITE} that replaces likelihood-based acquisition with a simple {\em proximity} heuristic over the current labeled pool. In EZR, the model maintains the same \emph{good} ($best$) and \emph{rest} partitions but selects the next point by preferring candidates closer to the current \emph{good} prototype than to the \emph{rest} prototype. When no such candidate is found in a small random sample, EZR falls back to random selection. This modification removes density estimation entirely and biases exploration toward regions already close to the best observed configurations. EZR is thus a non-parametric, distance-based acquisition strategy that avoids density estimation and focuses exploration on regions most similar to the current {\em good} frontier.
\color{black} 

\subsection{Statistical Methods}\label{stats}

All our algorithms are stochastic since they shuffle or mutate or select
rows at random.  Hence, to assess them, we 
must apply some statistical analysis
to the distributions seen in, say, 20 repeated trials  with different seeds.

We  ranked all our results using the  Scott-Knott \cite{Scott1974} recursive 
bi-clustering method. Results are sorted by median distance to heaven. \here{A11}\BLUE Scott-Knott operates on these distributions and is designed to consider run-to-run variability via its use of effect size and significance testing at each split.\color{black}
Scott-Knott splits the results where the expected difference in means is largest:
\begin{equation}
E(\Delta) = \frac{|l_1|}{l}(E(l_1)-E(l))^2 + 
\frac{|l_2|}{l}(E(l_2)-E(l))^2
\end{equation}
Here $|l_1|$, $|l_2|$ are split list sizes. If effect size and 
significance tests confirm differences, Scott-Knott ranks groups, 
recurses, and leaves rank 1,2,3...

Next, an effect size test uses Cliff's Delta~\cite{macbeth2011cliff}
to  check if the splits are sufficiently separate to be called ``different''. Cliff's scores range -1 (entirely less) to 1 
(entirely more), 0 means no difference. After that,
we apply a significance test  to check if the difference between two groups is more
that just accidental. For this, we use bootstrapping~\cite{EfroTibs93} to resample to estimate the  distributions, 
estimate confidence intervals, and check that if their differences
are just chance.   If the splits pass these effect size and significance tests,
Scott-Knott recurses over both halves. 

Scott-Knott is preferred over other tests such as (e.g.) Nemenyi \cite{nemenyi1963distribution} since it
(a)~avoids overlapping groups; (b)~tests for both significance and effect size;
(c)~handles overlapping distributions; and (d)~needs only $\log_2(N)$ 
comparisons.

\here{A12}\BLUE Since rankings reported by Scott-Knott already reflect both central tendency and variability across repeated trials and for space constraints, we report rankings and mean performance in this paper. Standard deviations are provided in an online appendix at \url{https://github.com/KKGanguly/NEO/blob/main/supplementary.pdf}.\color{black}

\subsection{Experimental Rig} 
 
All our algorithms begin by taking a data set and randomly shuffling the row order.
After that, our algorithms incrementally modify an initial set of selected rows.

Different algorithms start their optimization in different ways. For example,  LITE  starts with four randomly selected labeled rows while DEHB starts with an initially unlabeled population.

All our algorithms ran for a fixed budget (number of rows to label).
Initialization (described above) may consumes some init number of that budget. Labelling
budgets   were  initialized to a   small value (i.e. 6), then increased   while   increases lead to better performance.

\begin{tcolorbox}[colback=white,colframe=black,
title=\textbf{How the Summary Table is Constructed}, left=4pt,right=4pt, valign=top]
\label{box:construction}
\scriptsize
Table \ref{tab:percentbest_colrank} was constructed using   tables like the one  given below (e.g.).  Each cell in the full results table contains a $d2h$ score (from Equation~\ref{delta}-- so {\em lower} values are {\em better}) and its
Scott-Knott rank (e.g., ``45 b'' says that a $d2h$ of 45 was ranked ``b''; i.e. second place).  
The Scott-Kott letters let us color code the cells:
\raisebox{0.4ex}{\fcolorbox{black}{gold}{\rule{1em}{0em}}} best (a),
\raisebox{0.4ex}{\fcolorbox{black}{lightyellow}{\rule{1em}{0em}}} 2nd (b),
\raisebox{0.4ex}{\fcolorbox{black}{lightpurple}{\rule{1em}{0em}}} 3rd (c),
\raisebox{0.4ex}{\fcolorbox{black}{lightred}{\rule{1em}{0em}}} 4th (d),
\raisebox{0.4ex}{\fcolorbox{black}{lightgrey}{\rule{1em}{0em}}} 5th (e),
\raisebox{0.5ex}{\protect\fcolorbox{black}{lightblue}{\rule{1em}{0em}}} 6th (f),
\raisebox{0.5ex}{\protect\fcolorbox{black}{lightgreen}{\rule{1em}{0em}}} 7th (g),
\raisebox{0.5ex}{\protect\fcolorbox{black}{lightcyan}{\rule{1em}{0em}}} 8th (h),
\raisebox{0.5ex}{\protect\fcolorbox{black}{lightorange}{\rule{1em}{0em}}} 9th (i),
\raisebox{0.5ex}{\protect\fcolorbox{black}{lightpink}{\rule{1em}{0em}}} 10th (j).
Below are three example rows from the large table, showing how
the optimizer performances appear. Each column corresponds to an Optimizer with a specific budget (max labels), e.g., DEHB-6 is DEHB with budget as 6. Due to space constraints, we only show here a selected number of columns. For the full version
of this table, see \url{https://github.com/KKGanguly/NEO/blob/main/supplementary.pdf}.
\begin{center}
\scriptsize
\begin{tabular}{l|c|c|ccccccc}
Dataset &
$\Delta$ &
Before &
DEHB-6 & DEHB-12 & LITE-6 & RAND-6 & LINE-6 & SMAC-6 & TPE-6 \\ \hline
\textbf{player\_statistics} &
87 &
\cellcolor{lightpurple}38 c &
\cellcolor{lightyellow}15 b &
\cellcolor{lightyellow}14 b &
\cellcolor{gold}11 a &
\cellcolor{gold}9 a &
\cellcolor{gold}11 a &
\cellcolor{lightyellow}14 b &
\cellcolor{gold}9 a \\
\textbf{auto93} &
90 &
\cellcolor{lightred}56 d &
\cellcolor{lightpurple}38 c &
\cellcolor{lightyellow}35 b &
\cellcolor{gold}24 a &
\cellcolor{lightpurple}37 c &
\cellcolor{lightpurple}37 c &
\cellcolor{lightyellow}35 b &
\cellcolor{lightpurple}37 c \\
\textbf{Wine\_quality} &
78 &
\cellcolor{lightgrey}42 e &
\cellcolor{lightred}36 d &
\cellcolor{lightpurple}35 c &
\cellcolor{lightpurple}31 c &
\cellcolor{lightpurple}31 c &
\cellcolor{lightyellow}28 b &
\cellcolor{lightpurple}35 c &
\cellcolor{gold}23 a \\
\textbf{SS-G} &
91 &
\cellcolor{lightred}57 d &
\cellcolor{lightyellow}45 b &
\cellcolor{lightyellow}43 b &
\cellcolor{lightpurple}42 c &
\cellcolor{lightred}40 d &
\cellcolor{gold}35 a &
\cellcolor{lightyellow}37 b &
\cellcolor{lightgrey}33 e \\
\textbf{Health-ClosedPRs0006} &
98 &
\cellcolor{lightgreen}60 g &
\cellcolor{lightgreen}62 g &
\cellcolor{lightgrey}40 e &
\cellcolor{lightgrey}31 e &
\cellcolor{lightred}22 d &
\cellcolor{lightyellow}10 b &
\cellcolor{gold}2 a &
\cellcolor{lightgreen}57 g \\ \hline
\textbf{Percent best} &
-- &
0\% &
0\% &
0\% &
60\% &
40\% &
40\% &
40\% &
60\% \\
\end{tabular}
\end{center}
\paragraph{Step 1: Percent-best.}
For each optimizer and budget (e.g., DEHB-6, LITE-50, EZR-200),
we count how many datasets assign that optimizer rank ``a''.
If LITE-100 appears as ``x a'' in $n$ of the datasets,
its percent-best is:
\[
\text{percent-best} = \frac{n}{N_\text{datasets}} \times 100.
\]
\paragraph{Step 2: Normalized improvement ($\Delta \pm \text{Std}$).}
For each optimizer-budget pair, we compute the normalized improvement $\Delta$ (as defined in Equation~\eqref{delta}) across all datasets where data is available.
We report the mean $\Delta$ and its standard deviation across datasets.
For example, if LITE-6 achieves $\Delta$ values of 45 and 61 on two datasets, then:
\[
\Delta \pm \text{Std} = 53 \pm 8.
\]
Higher $\Delta$ indicates better relative improvement over the baseline.
\paragraph{Step 3: Column-wise ranking in the summary table.}
After computing percent-best and $\Delta \pm \text{Std}$ for each optimizer and budget:
\begin{itemize}[leftmargin=12pt]
\item Higher percent-best = better (colors apply).
\item Higher $\Delta$ = better relative improvement (no colors in $\Delta$ columns).
\item We assign colors to the top 3 optimizers per column in the percent-best section.
\end{itemize}
This process transforms thousands of per-dataset $d2h$ decisions
into the compact summary table shown in~Table \ref{tab:percentbest_colrank}.
\end{tcolorbox}
\begin{table*}[h!]
\centering
\caption{
Column-wise ranking for percent-best performance and normalized improvement ($\Delta$).
\raisebox{0.4ex}{\fcolorbox{black}{gold}{\rule{1em}{0em}}} best (a),
\raisebox{0.4ex}{\fcolorbox{black}{lightyellow}{\rule{1em}{0em}}} second (b),
\raisebox{0.4ex}{\fcolorbox{black}{lightpurple}{\rule{1em}{0em}}} third (c).
Higher percent-best and higher $\Delta$ (normalized improvement) is better.
For details on how this table was constructed see ``{\bf How the Summary Table is Constructed}'' (on this page).
}
\label{tab:percentbest_colrank}
\scriptsize
\renewcommand{\arraystretch}{1.3}
\setlength{\tabcolsep}{7pt}
\begin{tabular}{l|ccccccc|c@{~}|c@{~}|c@{~}|c@{~}|c@{~}|c@{~}|c}
  & \multicolumn{7}{c|}{\textbf{percent best}} & \multicolumn{7}{c}{\textbf{$\Delta \pm \text{Std}$}} \\
\textbf{budget $\Rightarrow$} 
& 6 & 12 & 18 & 24 & 50 & 100 & 200
& 6 & 12 & 18 & 24 & 50 & 100 & 200 \\
\hline
\textbf{LITE} 
& 32
& \cellcolor{gold}54
& \cellcolor{gold}70
& \cellcolor{gold}75
& \cellcolor{gold}82
& \cellcolor{lightyellow}88
& \cellcolor{gold}94
& 53±25
& 69±24
& 74±22
& 77±21
& 83±19
& 87±18
& 90±16 \\
\textbf{EZR} 
& \cellcolor{lightpurple}34
& \cellcolor{lightyellow}47
& \cellcolor{lightyellow}56
& \cellcolor{lightpurple}62
& \cellcolor{gold}82
& \cellcolor{gold}90
& \cellcolor{gold}94
& 51±25
& 65±24
& 71±23
& 74±22
& 82±19
& 88±16
& 90±14 \\
\textbf{LINE} 
& \cellcolor{gold}36
& \cellcolor{lightpurple}43
& \cellcolor{lightpurple}54
& 59
& \cellcolor{lightpurple}78
& \cellcolor{lightyellow}87
& \cellcolor{lightyellow}91
& 56±25
& 65±23
& 70±22
& 74±22
& 80±19
& 84±17
& 88±15 \\
\textbf{RANDOM} 
& \cellcolor{lightyellow}35
& \cellcolor{lightpurple}43
& 48
& 57
& 75
& \cellcolor{lightpurple}85
& \cellcolor{lightyellow}91
& 54±24
& 64±24
& 69±23
& 73±22
& 78±20
& 83±18
& 87±15 \\
\hline
\textbf{SMAC} 
& 28
& 38
& 49
& \cellcolor{lightyellow}63
& \cellcolor{lightyellow}79
& \cellcolor{lightpurple}86
& \cellcolor{lightpurple}90
& 50±26
& 61±26
& 67±24
& 70±28
& 79±22
& 85±18
& 86±21 \\
\textbf{TPE} 
& 27
& 41
& 51
& 55
& \cellcolor{lightyellow}79
& 83
& 83
& 46±33
& 63±29
& 63±32
& 67±32
& 77±24
& 81±22
& 82±24 \\
\textbf{DEHB} 
& 24
& 25
& 31
& 39
& 59
& 64
& 69
& 42±34
& 43±42
& 48±44
& 53±39
& 59±40
& 60±44
& 68±41 \\
\end{tabular}
\end{table*} 
\subsection{Results}\label{results}
\here{A13}
\BLUE
Table~\ref{tab:percentbest_colrank} shows results from 20 trials using budgets 
$\{6,12,18,24,50,100,200\}$. We stop at 200 since improvements 
rarely occur after 100 samples.
In the following, we compare the performance of Random, LINE, LITE, DEHB, SMAC, and TPE.

Table~\ref{tab:percentbest_colrank} reports $\Delta$ (Equation~\ref{delta}), 
the normalized improvement relative to the baseline before optimization.
All optimizers across all budgets show positive $\Delta$, demonstrating that 
{\em even our simpler optimization strategies find significant improvements.}
At budgets $\geq 100$, $\Delta$ approaches 90, indicating  {\em our optimizers find very large improvements.}

\begin{wrapfigure}{r}{2.8in} 
\begin{center} \includegraphics[width=2.5in]{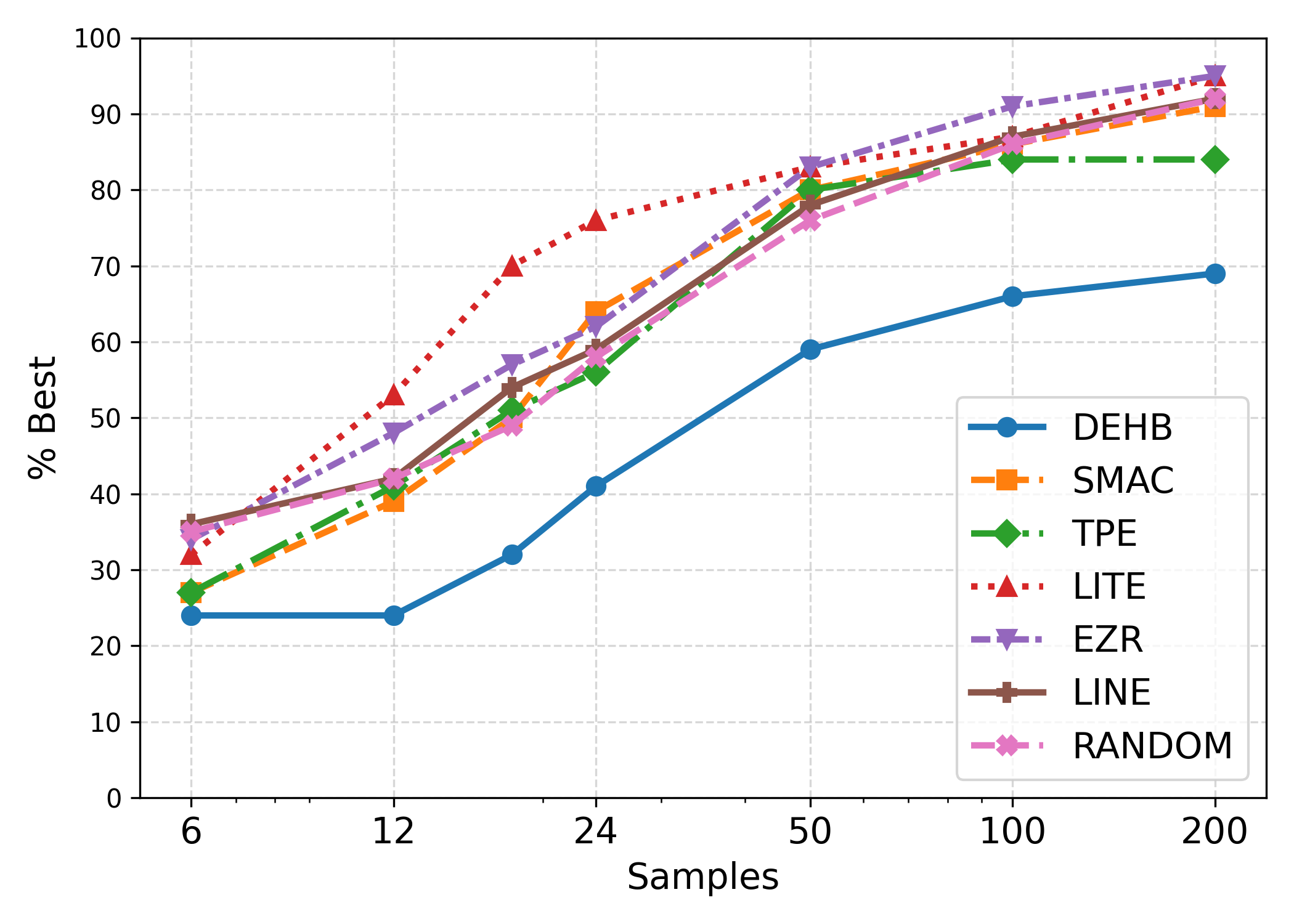}
\end{center}
\caption{For each sample size, in what percentage of the data sets, did a treatment achieve best-ranked performance? }\label{pcf} 
\end{wrapfigure}

The colored cells in each of the rows of
Table~\ref{tab:percentbest_colrank} show how often (in percentages) an optimizer achieved first-ranked results. Figure~\ref{pcf} summarizes those numbers. 

In Figure~\ref{pcf}:
\bi
\item In the zone we want to explore (the 100 rows or less required by 
\S\ref{hardlabel}), there  is diminishing returns after 100 samples.
\item
Random does surprisingly well but usually it does not perform best.
\item LITE performs the best considering all the budget levels. It is also evident that different optimizers work best for different sample sizes. Very small budgets favor simple, non-learning methods (LINE, RANDOM); intermediate budgets(12-50) favor LITE; and larger budgets show little separation between LITE and EZR.
\item In general, membership-query methods cannot bypass simple pool-based techniques within our label budgets. In all the budgets, at least one of the pool-based simple optimizer achieves the best rank. DEHB usually performs worse than anything else.
\ei \begin{wrapfigure}{r}{3.0in}
        \centering
        \includegraphics[width=\linewidth]{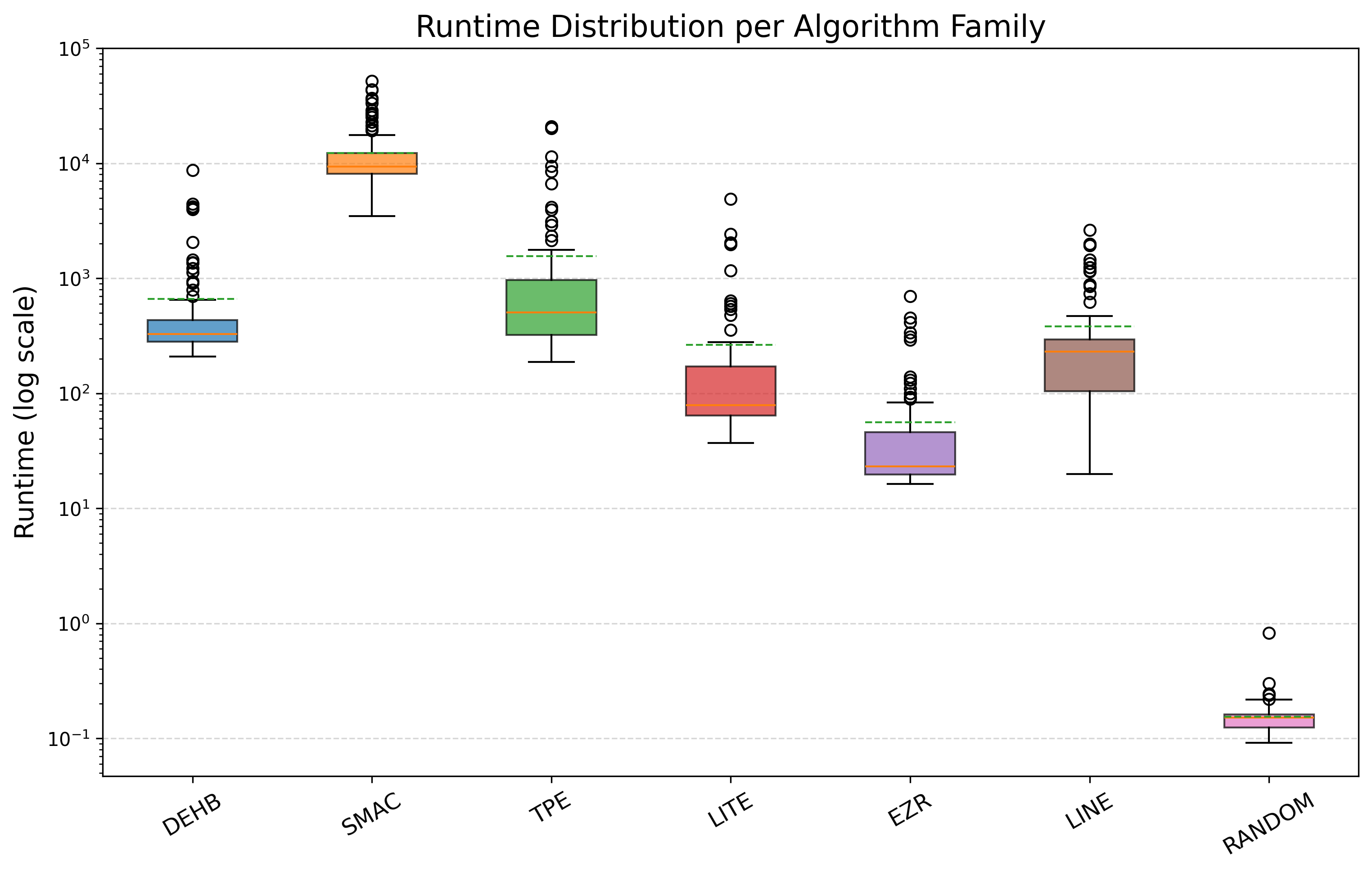}
        \caption{Runtime comparison (log scale, milliseconds) of different optimizers across datasets.}
        \label{fig:perf}
        \vspace{-0.3cm}
\end{wrapfigure}This results are  significant since 
our reading of the literature is that any newcomer who read that literature,
would be guided to relatively complex algorithms like DEHB and not
much simpler methods like LINE or LITE.

Figure~\ref{fig:perf} shows the runtimes of our different methods. Our
simplest pool-based method (RANDOM) that just samples the
buckets, runs $>$10,000 times faster than the membership query
ones. Out of the other two, EZR is the fastest followed by LITE and LINE. All Membership-query methods incur higher overhead. SMAC is the slowest where its runtime is more than two orders of magnitude higher than LITE and EZR. TPE is faster than SMAC but still over an order of magnitude slower than LITE and EZR. Although DEHB is the fastest of the membership query optimizers, its consistently poor performance in our setting makes it unsuitable for the scenarios considered in this study. 

The vertical axis of Figure~\ref{fig:perf} is in milliseconds, so it could
be argued that, in terms of absolute values, these are all similar.
We would dispute that, saying that our evaluation times
here are near instant (see the labels). In practice,
each evaluation could be very slow. For example, to check
a proposed configuration for a Makefile, we may need to
recompile an entire system, then run an extensive test suite. Moreover, runtime difference compounds across hundreds of configurations, repeated stochastic trials, multiple datasets. Thus, magnitudes of millisecond-level gaps translate into minutes or hours at scale which would become of great practical importance. Finally, simple optimizers considered in our study reach strong performance earlier which enables early stopping. For example, EZR can often be stopped at 100 labels, whereas SMAC may require 200 labels to reach comparable performance (Table \ref{tab:percentbest_colrank}), effectively doubling the total runtime. Together, these effects make simpler methods substantially more efficient in practice.

In summary:
\begin{conclusion}{Core result:}
For the NEO problems explored here, very simple methods run much faster and find  better optimizations as  more complex methods (especially when 
the labeling budget is very tight: say, 4 dozen labels or less).
\end{conclusion}
\color{black}
\section{When Simplicity Might Fail} \label{bingo_eff}
This section asks when we should mistrust the simplicity results of the last section.
This section reports a quick way to peek at data to  recognize if it is amenable to very simple reasoning.

While exploring the above simplicities, we looked into the topology of our data.
We found previously unreported phenomenon— which suggests that for optimization in SE, CPU-intesive methods may     be unnecessary:
\begin{quote}{\em The {\bf BINGO} effect:
When SE data is split into $n$ buckets across $d$
dimensions divided into $b$ bins, most data goes to surprisingly few
buckets; i.e. $n \ll b^d$.}
\end{quote}
For  example, in one  study with 10,000 rows, we had expected to see
expect $b^d=4096$ buckets from $d=4$ dimensions split into $b=8$ bins.
Yet in practice, we only  found $100$ used buckets. Labeling millions of rows is clearly unnecessary when only (say)
100 distinct scenarios exist. 

\begin{wrapfigure}{r}{3.5in}
\vspace{-0.5cm}
 \caption{Data can be divided by adding $bins$ per dimension, or adding dimensions $d$. On the left, increasing  data  divisions  increases the number
 of   buckets with data. On the right, it does not (since the data is more clumped).}\label{eg1234}
 \begin{center}
 \includegraphics[width=3.5in]{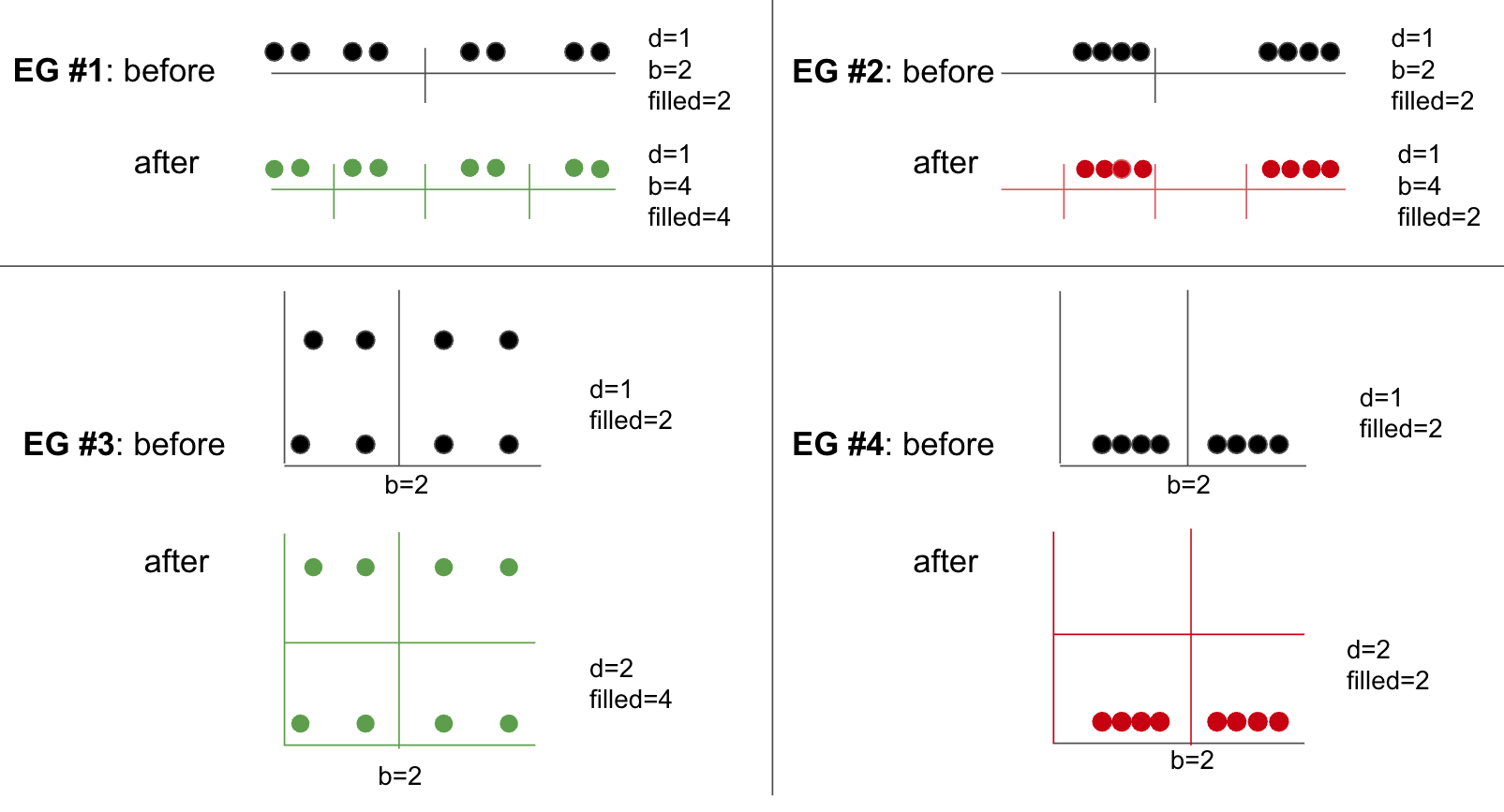}
 \end{center}
 \end{wrapfigure}
The rest of this section documents this  BINGO effect. 
To understand  BINGO's concept of ``buckets'',   consider $d$ \textbf{dimensions} (features) split 
into $b$ \textbf{bins} (ranges) to form $b^d$ \textbf{buckets}. Data rows are assigned to buckets  according to  where they call along the   dimensions.
For this $b^d$ space,
Figure~\ref{eg1234} offers examples  where  increasing the number of data divisions results
in empty buckets.

EG~\#1 in Figure~\ref{eg1234} shows  that
if   one-dimensional data is spread out, then   more buckets  can be filled if we  increase the number of bins from (e.g.) two to four.

On the other hand, as shown in EG~\#2, if the data clumps together, then doubling the number  of bins can lead to  empty buckets.
In this example, increasing the number of bins from two to four leads
to 50\% empty buckets.

Similarly, if two-dimensional  data is spread out, adding dimensions fills more buckets. As seen in EG~\#3, doubling the dimensions can lead to double the number of filled buckets.

On the other hand, as seen in EG~\#4,  if two-dimensional data clumps, then adding new dimensions may yield empty buckets.

Figure~\ref{eg1234} illustrates the {\sffamily BINGO} effect: in clumped data,
adding divisions does not increase occupied regions. Algorithm~1 
tests for this effect by counting filled buckets as divisions increase, selecting
dimensions via Table~\ref{bingrow}. To filter noise and outliers, buckets require
a minimum capacity ({\em minPts}). Following DBSCAN guidelines~\cite{Schubert2017},
line 10 of Algorithm~1 
requires filled buckets to contain at least $2d$
{\em minPts} rows.

 
\begin{table} 
\caption{Finding Multi-Dimensional Buckets.}\label{bingrow}
\footnotesize\begin{tabular}{|p{.95\linewidth}|}\hline
Data often lies on simpler, lower-dimensional "manifolds" 
\cite{zhu2005semi}. Dimensions should reflect underlying manifold 
shape. 
Pearson's 1901 {\bf PCA} \cite{pearson1901principal} seeks data's 
overall shape via eigenvectors (max variance directions) from 
covariance matrices.  In PCA, data is projected to lower dimensions using 
the top 
eigenvectors. PCA is costly for large datasets and does not work
for mixtures of numeric and symbolic columns. \\\\
\rowcolor{blue!10}

Nyström alogirthms approximates eigen decomposition for large kernel matrices.
{\bf Fastmap} \cite{faloutsos1995fastmap}, 
a Nyström-like method that derives components from distant "corners" ($k$).
In operation, it picks a random row $r$; then selects row $k_1$ far from $r$.
We call these $k$ rows, the ``korners''.
In the selects a row  
$k_2$ far from $k_1$. The line $k_1-k_2$ captures max data spread 
(like PCA's first component).\\\\
Row $i$ can then be map to $b$ bins via the cosine rule:
\[
x=\text{int}\left(\frac{D(k_1,i)^2 + D(k_1,k_2)^2 - D(k_2,i)^2}
{2 \cdot D(k_1,k_2)^2 \cdot b}\right)
\]
\\\rowcolor{blue!10}

This maps rows to $b$ bins along one dimension defined by korners 
$k_1,k_2$. Multiple dimensions need more korners ($k_3, \dots$), 
each maximally distant from prior corners. Rows map to bins on 
each dimension.\\\rowcolor{blue!10}

For mixed data (numbers/symbols), Fastmap uses {\bf Aha's distance} 
\cite{aha1991instance}: Numbers: $\mathit{dist}(x,y) = |x-y|$ 
(normalized 0..1). Symbols: $\mathit{dist}(x,y) = 0$ if $x=y$, 
else $1$. Missing: If both missing, $\mathit{dist}(x,y)=1$. If 
only $x$ missing, $x$ is $0$ if $y>0.5$, else $1$.
In this work, $D(x,y) = \left(\sqrt{\frac{\sum_{i=1}^N \mathit{dist}(x_i,y_i)^2}{N}}\right)/\sqrt{N}$; 
$0\le D \le 1$.\\\hline
\end{tabular} 
\end{table}

 
\begin{algorithm}
 \footnotesize
    \caption{Bucket Building with Randomized Parameters}
    \label{bingrows}
    \begin{flushleft}
    \textbf{Input:} Table~\ref{combinedtable} - a list of datasets with rows and columns \\
\hspace*{1em} $\mathit{Buckets(data, d, b)}$ — divide on $b$ bins, $d$ dims, see Table~\ref{bingrow}. \\
\hspace*{1em} $\mathit{RandomInteger(a, b)}$ — returns random integer in $[a,b]$ \\
\hspace*{1em} $\mathit{RandomChoice(Table~1)}$ — selects a random dataset \\
    \textbf{Output:} Two bucket sets $n_1$ and $n_2$ with sufficient row counts 
    \\
    1. $count \gets 0$ \\
    2. \textbf{While} $count < 1000$ \textbf{do} -- try this, many times\\
    3. \hspace*{1em} $d_1 \gets RandomInteger(3..8)$\\
    4. \hspace*{1em} $d_2 \gets RandomInteger(d_1..8)$\\
    4. \hspace*{1em}  $b_1 \gets RandomInteger(3..10)$ \\
    5. \hspace*{1em} $b_2 \gets RandomInteger(b_1..10)$ \\
    6. \hspace*{1em} \textbf{if} $d_2 > d_1$ \textbf{or} $b_2 > b_1$ \textbf{then} -— data divisions have been found\\
    7. \hspace*{2em} $count \gets count + 1$ \\
    8. \hspace*{2em} $data \gets \mathit{RandomChoice}(\text{from Table 1})$ \\
    9. \hspace*{2em} \textbf{for} $(d, b) \in \{(d_1, b_1),~(d_2, b_2)\}$ \textbf{do} \\
    10. \hspace*{3em} $n \gets \{x \in$ buckets$(data, d, b)~|~\text{len}(x) \ge 2d\}$ \\
    11. \hspace*{3em} Append $|n|$ to result list \\
    12. \hspace*{2em} end for\\
    13. \hspace*{2em} \textbf{print} $\left(\lfloor \log_2(|\text{data.rows}|) \rfloor,~\text{result}[0],~\text{result}[1] \right)$ \\
    14. \hspace*{1em} \textbf{end if}\\
    15. \textbf{end while}
    \end{flushleft}
\end{algorithm}

 \begin{figure} 
\caption{ Algorithm~1, applied to Table~\ref{combinedtable}. {\em Buckets (after)} counts the  filled buckets seen {\em after}  increasing the data divisions.   Y-axis shows mean results for data sets grouped  by $\log_2$ of the number of rows. Blue lines indicate dataset sizes. Red lines show the bucket counts from randomly chosen bins/dimensions. Yellow lines show buckets after using more bins/dimensions.  }\label{bingos}
\includegraphics[width=3.5in]{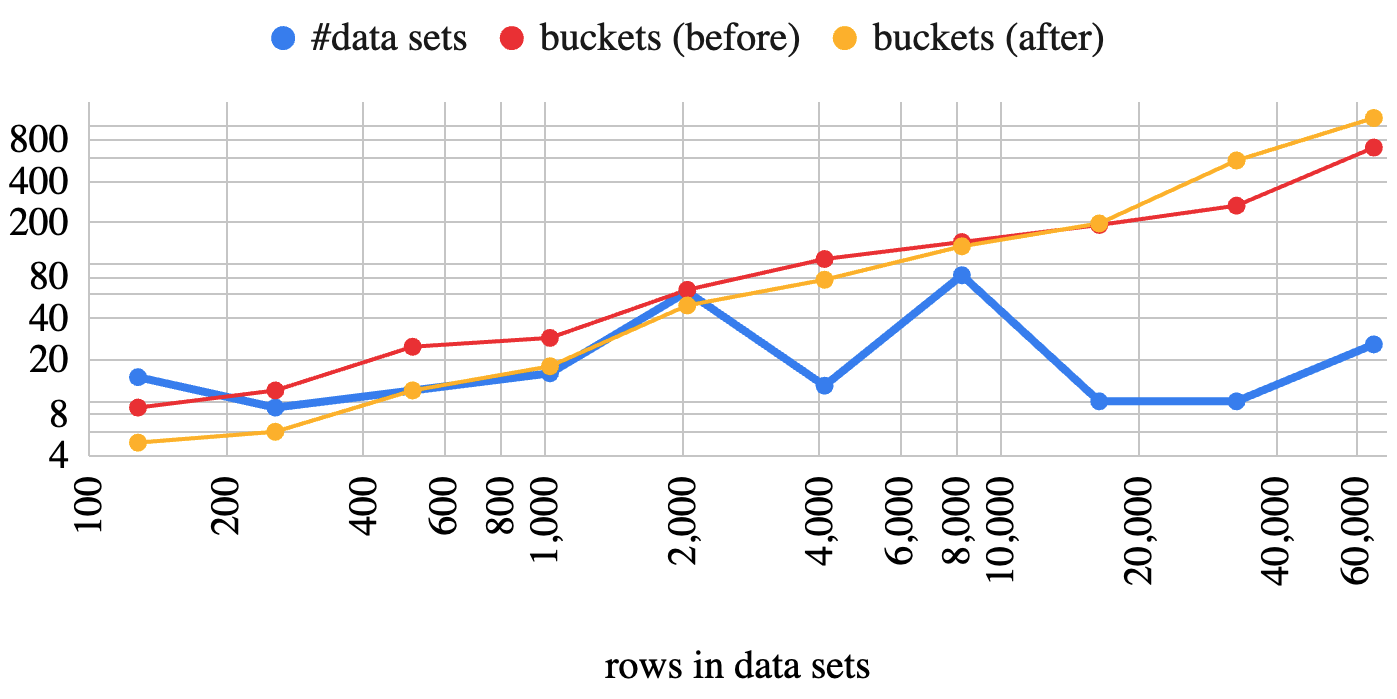}
\vspace{-0.5cm}
\end{figure}
  Figure~\ref{bingos} shows the output of Algorithm~1  applied to   Table~\ref{combinedtable}.  The blue lines shows that most of our data falls in the range 1,000 to 10,000 rows.
The red lines shows the   buckets   generated using some number of   bins and dimensions. The yellow line shows     buckets seen after
trying to use more bins and/or more dimensions.

In Figure~\ref{bingos},
most buckets are empty (which is the {\sffamily BINGO} effect). Given the ranges of $b$ bins and $d$ dimensions explored in Algorithm~1,
the mean $b^d$ is around 5000. Yet as shown in this figure, data compresses   to around 100 buckets (see the median $y$ values).

The ratio of empty buckets increased with the number of rows in a data set:
\begin{itemize}
\item
 256 rows go to 10 buckets $(10/256\approx 4\%) $
\item  32,000  rows go to 500 buckets
$(500/32,000\approx 1\%)$.
\item  64,000 go to around
1\% of the possible buckets.  
\end{itemize}
For most of our samples, the yellow line is below the red line. That is, increasing the number of data divisions does not increase the number of filled buckets.
To say that   another way,
our data is not spread evenly across the search space. Rather, it clumps
into just a few corners.

We say that   BINGO  tells us when simpler methods are not appropriate.
\here{A14}\BLUE
To make the BINGO effect actionable, we define a simple \textit{compression ratio} that measures how strongly the data collapses into a small number of regions. Let $N$ be the number of rows and let $n$ be the number of buckets occupied by the data. The \textit{compression ratio} is defined below.
\[
r = \frac{n}{N}.
\]

Empirically, when $r < 0.05$, the data shows a strong BINGO effect as most configurations fall into very few distinct regions. In this case, simple pool-based methods typically achieve strong performance using no more than 50 labeled examples. When $0.05 \le r < 0.20$, the BINGO effect is moderate: lightweight methods can still be effective, but may require larger budgets (50-100 labels). When $r \ge 0.20$, the data is more evenly spread across the space, and simple methods are likely to fail. More complex optimizers should be preferred in this case.

These thresholds do not guarantee success. However, they offer a low-cost test for identifying when data-light optimization is likely to succeed and when it should be avoided.
\color{black}
\begin{conclusion}{When to be simple:}
If data collapses to
many thousands of bins, or more, then reasoning will need more than the small number of samples used in \S\ref{expr}.  
\end{conclusion}


 \section{Discussion}
 \subsection{Threats  to Validity}\label{tov}

 
{\em Evaluation bias:} This paper   measure   success with {\em distance to heaven}. We have repeated the above analysis using another evaluation bias:
the Chebyshev distance to utopia measured preferred by MOEA/D researchers~\cite{zhang2007moea}. While this changed some of the numbers seen above,
the main conclusions persisted; i.e. very simple methods that used very few labels
performed as well (or better) than using more data and/or more complex algorithms.

{\em Order bias:} As with any stochastic analysis, 
the results can be effected by the ordering in which
the examples are randomly selected.   To mitigate this order bias, we run
the experiment 20 times, each time using a different seed for the random number generator.

{\em Learner bias:} This paper has compared pool-based inference to member-ship query methods. The case for that comparison was made above but, as with any empirical study, some new algorithm could arrive tomorrow that changes all these results. For example, there exists one potentially better successor to DEHB, called  PriorBand~\cite{Mallik23}. While the underlying principles of  PriorBand are  applicable to many optimization problems,  PriorBand is specifically tailored for deep learning applications (which are not studied here). Hence, for this study, we chose to remain with DEHB. 

In our view, {\em sampling bias} is the biggest threat to the validity of this paper. 
Our reading of this literature is that most papers evaluate themselves with five data sets or  less (sometimes, only one) so the sample size used here ($N=39$) is far larger than most other papers. That said, this paper does not explore
data from the test case generation or test case reduction literature. This
is clearly a direction for future work. 

More generally, another complaint about our sample
might be that  our datasets are {trivially small} \cite{yang2022survey},
especially  when compared to large SE textual/image/code data sets. In reply, we say our
Figure~\ref{combinedtable} datasets were used in top SE journals
like IEEE Trans. SE~\cite{peng2023veer, Chen19, nair2018finding}, Empirical
Softw. Eng.~\cite{peng2023veer, xia2020sequential,guo2018data}, and
ACM Trans. SE Methodologies~\cite{lustosa2024learning}, thus demonstrating
that  our research community finds this kind of data interesting.

Another complaint is that our datasets are
{simplistic old-fashioned tabular} data. We argue tabular data is not
simple or outdated. Its processing is challenging, even for LLMs.
Somvanshi et al.~\cite{somvanshi2024survey} report that ``despite
deep learning’s success in image and text domains, tree-based
models like XGBoost and Random Forests continue to outperform neural
networks on medium-sized tabular datasets. This performance gap
persists even after extensive hyperparameter
tuning''~\cite{somvanshi2024survey}.

Tabular data is widely used. Somvanshi
et al.~\cite{somvanshi2024survey} call it ``the most commonly used
data format in many industries...finance, and transportation.''.
Commercial use
in data synthesis/privatization for health/government~\cite{Ling24, menzies2023best}. GitHub-scale
code analytics uses tabular data (e.g.,
CommitGuru\footnote{\url{http://commit.guru/}}, used in many SE
papers\footnote{See \url{http://tiny.cc/guruSince2019} for
CommitGuru usage.}). Software HPO uses data like
Table~\ref{combinedtable}; many tasks involve cloud software
HPO~\cite{peng2023veer}. Xia et al.~\cite{xia2020sequential}
analyzed 1600 GitHub projects using tables (13 indicators x 60
months). Non-SE HPO techniques
(Hyperband~\cite{li2018hyperband},
SMAC3~\cite{lindauer2022smac3}, DEHB~\cite{awadijcai2021p296})
also use tabular benchmarks.

\here{A15}
\BLUE
Real-world SE labels are often expensive, slow, noisy (e.g., flaky tests, imperfect logs, heuristic labeling). In this study, we intentionally use pre-labeled datasets to isolate the effect of the distributional structure of the configuration–outcome space on label efficiency under controlled and reproducible conditions, similar to other studies in this domain \cite{nair2018finding,chen2026promisetune,van2017automatic,chen2021efficient}. While noisy labels can affect performance and robustness \cite{yu2019improving}, systematically studying noise requires careful modeling choices such noise type, magnitude, and correlation that are beyond the scope of this work. Investigating the interaction between BINGO-style collapse and noisy labels is an important direction for future work.
 \color{black}
 \subsection{Future Work}
 As to future work, we should  extend the    {data-light challenge} to other SE domains to see 
{\em when will a handful of labels suffice for SE tasks?} As shown here, exploring that question can lead to some surprising results that can dramatically simplify
the process of reasoning about software.

\here{A16}
\BLUE Also, future work could check the value of our lightweight methods for generative AI workflows. Recent work on prompt optimization shows that large prompt search spaces often reduce to a minimized set of influential tokens or clusters, with many prompts exhibiting redundant or near-equivalent behavior \cite{zhou-etal-2023-survival}. Determining whether such redundancy induces effective search-space collapse and whether data-light methods can exploit it to guide early prompt selection remains an open research question.

Further, there is much current interest in Agentic systems. Choosing between different task breakdowns and then learning the control parameters for each part of that architecture are two tasks
well represented in the MOOT repository. Hence we say our lightweight methods could be very useful for quickly designing Agentic systems. \color{black}

Furthermore, we need more work on   the BINGO effect. Perhaps the maths \S\ref{maths}  is overly pessimistic about Reward-free reasoning. If real-world data always collapses to just tiny corners of the possible search space, that would make the
verification of AI software  easier. This would have a major impact on (e.g.) the safety
analysis of   autonomous software for  cars.

\section{Conclusions}
This paper has argued that, for certain SE problems, complex AI methods
can and should be avoided.   
This position runs
against the current fashion. Much of the current literature begins not with the
problem, but with the tool; i.e. ``here is an LLM; now what can I do with
it?''.
We caution against that approach, since at least for some SE tasks since,
  data labeling shortages mean  we may have access to very little   ``gold standard'' labeled data.  

 Fortunately, the class of SE optimization tasks studied here require very few labels. For NEO problems (Near-Enough Optimization) with labeling budgets \{6, 12, 18, 24, 50, 100, 200\}, gains after 100 labels were rare. Table~\ref{tab:percentbest_colrank} shows our methods reached maximum $95\%$ of optimum, with simple methods (LINE/LITE) consistently outperforming heavier baselines like DEHB. Figure~\ref{fig:perf} demonstrates these simpler methods are faster, offering strong quality/runtime trade-offs with very few labels.
 
We explain the curious and surprising   success of our data-light methods as follows:\bi
\item As discussed in \S\ref{cog}, humans have a cognitive  bias
that blocks their exploration of simpler solutions. The simpler methods found here have  not been previously reported since 
{\em other researchers were not looking for them}. 
\item
 As discussed in \S\ref{maths}, there is math suggesting that, for certain tasks (i.e. Near Enough optimization), we should expect that a few dozen labels will suffice.
\item
As shown  in \S\ref{bingo_eff}, there are certain topological features of data that make it amenable to reasoning with only a few labels.
Here, we are referring to the BINGO effect where most of the data clustered into a few small number of bins. 
As mentioned above, when data collapses to just a few  buckets, it does not need to be sampled with 1,000,000 probes.
\ei

\BLUE
\here{A17}
Finally, for practitioners, we close with
practical rules on when {\em not}
to use simple method like those
explored here:
\bi
\item
{\bf Rule 1:} Do not use our simple methods for mission-critical/safety-critical applications (when tasks require precision beyond 90–95\% of optimal)
\item
{\bf Rule 2}: Do not our simple
methods if the data landscape is ``too complex''. To operationalize this rule, run Algorithm~1 and compute the compression ratio $r = n/N$ (occupied buckets / rows).
After that:
\bi
\item
$r < 0.05$ ({\bf Strong BINGO effect}): Simple methods   using
$r \le 
50$ labels should  suffice;
\item
$0.05 <= r < 0.20$ ({\bf Moderate BINGO effect}): Use 50–100 labels.
\item
$r >= 0.20$  ({\bf No BINGO}): Data is too fragmented for NEO; use sophisticated optimizers or larger budgets
(additionally, avoid simple methods when buckets grow linearly with new sub-divisions of the dimensions or increase in the number of bins).
\ei
\ei
 \color{black}

\section*{Data Availability}
All our scripts  are online at \url{https://github.com/KKGanguly/NEO}. Also, for test data, see the ``optimize'' directory of Chen and Menzies' MOOT repository at \url{http://github.com/timm/moot}.
 
\bibliographystyle{IEEEtran}
\bibliography{_sorted}

\end{document}